\documentclass[twocolumn]{aastex62}
\pdfoutput=1

\usepackage{appendix}
\usepackage{epstopdf}

\shorttitle{High-Resolution ALMA Images of WL 17}
\shortauthors{Gulick et al.}

\usepackage[T1]{fontenc}
\usepackage{amsmath}
\usepackage{gensymb}
\usepackage{booktabs}
\usepackage{dcolumn}
\usepackage{lineno}
\usepackage{multirow}

\begin{document}

\title{Detection of Substructures in Young Transition Disk WL 17}

\correspondingauthor{Hannah C. Gulick}
\email{hannah\_gulick@berkeley.edu}

\author[0000-0002-0786-7307]{Hannah C. Gulick}
\affiliation{Center for Astrophysics | Harvard \& Smithsonian  \\
Cambridge, MA 02138, USA}
\affiliation{University of California, Berkeley \\
Berkeley, CA 94703, USA}

\author{Sarah Sadavoy}
\affiliation{Department of Physics, Engineering Physics, and Astronomy, Queen's University, \\
Kingston, ON K7L 3N6, Canada}

\author{Luca Matr\`{a}}
\affiliation{Centre for Astronomy, School of Physics, National University of Ireland Galway, University Road, Galway, Ireland}
 
\author{Patrick Sheehan}
\affiliation{National Radio Astronomy Observatory,\\
Charlottesville, VA 22901, USA}

\author{Nienke van der Marel}
\affiliation{University of Victoria,\\
Victoria, BC V8P 5C2, Canada}

\begin{abstract}

WL 17 is a young transition disk in the Ophiuchus L1688 molecular cloud complex. Even though WL 17 is among the brightest disks in L1688 and massive enough to expect dust self-scattering, it was undetected in polarization down to ALMA's instrument sensitivity limit.  Such low polarization fractions could indicate unresolved polarization within the beam or optically thin dust emission.  We test the latter case by combining the high sensitivity 233 GHz Stokes I data from the polarization observations with previous ALMA data at 345 GHz and 100 GHz. We use simple geometric shapes to fit the observed disk visibilities in each band. Using our simple models and assumed dust temperature profiles, we estimate the optical depth in all three bands.  The optical depth at 233 GHz peaks at $\tau_{233} \sim 0.3$, which suggests the dust emission may not be optically thick enough for dust self-scattering to be efficient. We also find the higher sensitivity 233 GHz data show substructure in the disk for the first time. The substructure appears as brighter lobes along the major axis, on either side of the star. We attempt to fit the lobes with a simple geometric model, but they are unresolved in the 233 GHz data. We propose that the disk may be flared at 1 mm such that there is a higher column of dust along the major axis than the minor axis when viewed at an inclination. These observations highlight the strength of high sensitivity continuum data from dust polarization observations to study disk structures.

\end{abstract}


\section{Introduction} \label{sec:intro}

WL 17 is located in the L1688 region of the Ophiuchus molecular cloud 137 pc away \citep{sheehan, ortiz}. Also known by Oph-emb-20 \citep{2009ApJ...692..973E} and GY 205 \citep{1992ApJ...395..516G}, this young stellar object (YSO) has a known disk and appears to be  embedded in an envelope of material from its natal dense core. Therefore, its emission peaks in the mid-infrared and is highly extincted at optical wavelengths \citep{evans, 2009A&A...498..167V, 2010ApJS..188...75M, van der marel}.

A previous study by \citet{sheehan} revealed a central cavity in WL 17's disk with a radius of $\sim$13 AU. Cavities are generally detected in Class II disks \citep{andrews,francis}, and have been attributed to physical processes like planets carving gaps \citep[e.g.][]{dodson}, photoevaporation \citep[e.g.][]{alexander}, dust grain growth \citep[e.g.][]{dullemond}, or stellar and disk winds \citep[e.g.][]{wind}. As these processes take time to clear out a central hole, it is unsurprising that cavities are rarely found in disks that are still embedded, like WL 17. Recent high-resolution millimeter observations with the Atacama Large Millimeter/submillimeter Array (ALMA) are beginning to report more detections of gaps and cavities in younger protostellar disks \citep{segura2020, sheehan2020}.

In a recent dust polarization survey with ALMA of all the embedded YSOs in the Ophiuchus molecular cloud, \citet{sarah} found that WL 17 was completely undetected in polarization at 1.3 mm with a 3$\sigma$ upper limit of <0.3$\%$, which is within the instrument noise. WL 17 is the brightest source in the sample to be undetected in polarization, even though it is brighter and more massive than other disks that exhibited polarization.

Of the disks that are detected in polarization, the majority have signatures consistent with dust self-scattering processes e.g. \citep{bacciotti, dent, sarah}. Polarization from self-scattering is most efficient when the dust grains have a maximum size of $\sim\lambda/2\pi$, where $\lambda$ is the observing wavelength \citep{2015ApJ...809...78K, lin}, and the emission is optically thick \citep{2017MNRAS.472..373Y}. The lack of detection in WL 17 could indicate non-optimal grain sizes, low optical depth, or unresolved polarization structure \citep{2015ApJ...809...78K, 2016ApJ...820...54K, 2017MNRAS.472..373Y, lin, hull19, tazaki, guillet}.

In this paper, we model the intensity profile of WL 17 to measure the dust optical depth and better characterize the lack of polarization in this disk.  We combine high sensitivity Stokes I continuum data from the polarization survey of \citet{sarah} at 233 GHz with archival data at 345 GHz and 100 GHz from \citet{sheehan}. With the high sensitivity continuum from the polarization observations, we also detect a double-lobed substructure in the WL 17 disk for the first time.

This paper is organized with Section \ref{sec:obs} describing our observations. Section \ref{sec:analysis} describes the geometric disk models and the double-lobed feature. In Section \ref{sec:discuss}, we discuss the features and measure the optical depth. Finally, we summarize the paper in Section \ref{sec:conc}.

\section{ALMA Observations} \label{sec:obs}

WL 17 was observed by ALMA in Bands 3, 6, and 7 as a part of three different projects. Table \ref{tab:obs} summarizes the observation logs for the three Bands. At 233 GHz ALMA was used in full polarization as part of a larger survey \citep{sarah}. The observations were taken over three blocks of 3 to 4.5 hours to have sufficient parallactic angle coverage for calibration. The correlator was set to standard full polarization for Band 6 with each baseband at approximately 2 GHz bandwidth and 64 channels. For observations at 100 GHz, all four basebands were configured for continuum observations, with approximately 2 GHz of continuum bandwidth per baseband. For observations at 345 GHz, two of the four basebands were configured for continuum observations with a total of approximately 3.75 GHz of continuum bandwidth.

The 233 GHz data were calibrated using the standard pipeline for polarization data and then self-calibrated with two rounds of phase-only self calibration. See \citet{sarah} for more details on the observations. The 100 GHz and 345 GHz data were calibrated following \citet{sheehan}.


\begin{deluxetable*}{cccccc}
\tablecaption{Observation Summary of WL 17 \label{tab:obs}}
\tablehead{
\colhead{Frequency} &  \colhead{Project ID} &\colhead{Date Observed} & \colhead{Baselines} & \colhead{Time on Source} & \colhead{Bandwidth} \\
\colhead{(GHz)} &  \colhead{} &\colhead{} & \colhead{(m)} & \colhead{(s)} & \colhead{(GHz)}
}
\startdata
100 & 2015.1.00761.S & 2015-10-31, 2016-04-17 & $14-1530$  & 400 & 7.5 \\
233 & 2015.1.01112.S & 2017-05-20, 2017-07-11, 2017-07-14 & $15-2647$  & 432 & 7.5 \\
345 & 2015.1.00761.S & 2016-05-19, 2016-09-10  & $15-3140$ & 90  & 3.75
\enddata
\end{deluxetable*}

\begin{figure*}[th!]
\includegraphics[width=175mm]{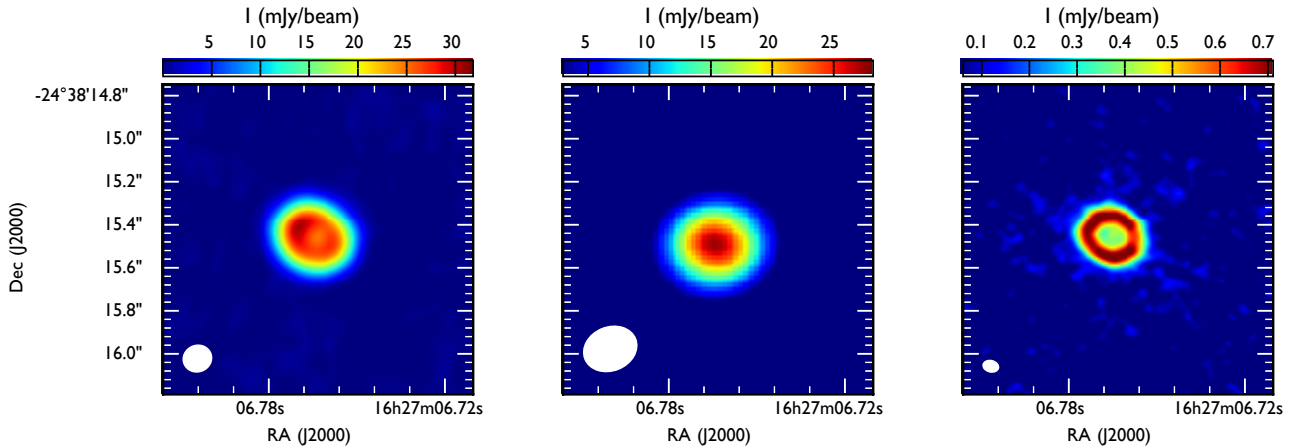}
\caption{ The 345 GHz (left), 233 GHz (middle), and 100 GHz (right) data cleaned with Briggs weighting and a robust of 0.5. See Table \ref{tab:maps} for the beam sizes.
\label{fig:rp5}}
\end{figure*}

Figure \ref{fig:rp5} shows an image of WL 17 at each wavelength using the CASA function \textit{clean} interactively and Briggs weighting with a robust parameter of 0.5. Table \ref{tab:maps} lists the sensitivities, resolutions, and signal-to-noise ratio (SNR) for each band with a robust value of 0.5.  We also list the sensitivity, resolution, and SNR for robust=-2 at 233 GHz and 345 GHz, which we use in Sections 3 and 4. Note that due to the difference in resolution, the SNR will vary with sensitivity and beam size as well as source properties. To better compare the SNRs, we smoothed the 100 and 345 data to the same resolution as the 233 GHz data and compared the point mass sensitivity using dust opacity from \citet{andrews2009} and assumed optically thin emission at a temperature of 30 K. The mass sensitivity of the 233 GHz data is 6 times better than the sensitivity in the smoothed 100 GHz data and 8 times better than the sensitivity in the smoothed 345 GHz data.


\begin{deluxetable}{cccccc}[h!]
\vspace{5mm}
\tablecaption{Map Sensitivities and Resolutions \label{tab:maps}}
\tablehead{
\colhead{Band} &  \colhead{Robust} &\colhead{Resolution} & \colhead{PA} & \colhead{RMS} & \colhead{SNR} \\
\colhead{(GHz)} &  \colhead{} &\colhead{(")} & \colhead{(\degree)} & \colhead{($\mu$Jy/beam)}
}
\startdata
345 & 0.5 & 0.14 x 0.13 & -68.5 & 270 & 120\\
345 & -2.0 & 0.12 x 0.12 & -7.4 & 730 & 50\\
233 & 0.5 &  0.26 x 0.21  & -68.2 &  35 & 800\\
233 & -2.0 & 0.22 x 0.14  & -63.0 & 63 &  450\\
100 & 0.5 & 0.08 x 0.06 & 76.5 & 35 & 20 \\
\enddata
\tablecomments{The resolution and sensitivity for images cleaned with a robust = 0.5 or robust = -2 for each frequency. The 100 GHz image is only shown for a robust of 0.5 as the ring structure is below 5$\sigma$ when a robust = -2 is used. }
\end{deluxetable}

\section{Image Analysis} \label{sec:analysis}

The multi-frequency data we have for WL 17 (see Figure \ref{fig:rp5}) have different resolutions and the SNR is much higher for the 233 GHz data than at the other two wavelengths (see Table \ref{tab:maps}), which makes a direct comparison between wavelengths challenging. There is also evidence of structure in the disk.  The 100 GHz data appears to show an asymmetric, horseshoe-like structure, although the difference between the eastern part of the ring and the dip in the western part is less than 3$\sigma$ and should not be over interpreted. The 345 GHz image also appears to be asymmetric, with a significantly brighter blob in the north-east by roughly 9$\sigma$ compared to the south-west part of the ring. We discuss possible sources of asymmetry in Section \ref{sec:dlemission}. The 233 GHz image has too low spatial resolution to identify possible asymmetries. Since the asymmetries detected at 100 GHz and 345 GHz are relatively minor compared to the peak brightness of the ring, we will model the disk assuming uniform flux distribution.

\subsection{Modeling the Structured Disk} \label{sec:Gaussian}

Since WL 17 has a known cavity \citep{sheehan}, we approximated the disk as a modified Gaussian ring model: \\

\begin{equation}
I(r) = f_p \exp{\left\{-0.5 \left( \frac{r-r_0}{\sigma}\right)^2\right\}}
\end{equation}

\noindent where $I$ is the radial surface brightness of the disk, $f_p$ is the peak intensity, $r_0$ is the cavity radius, and $\sigma$ is the standard deviation of the Gaussian. Our model also includes four geometric parameters: the inclination $i$, position angle $\phi$, and offset coordinates $d\alpha$ and $d\delta$ from the phase center.

We used a Monte Carlo Markov Chain (MCMC) fitting algorithm utilizing the Python packages \textbf{galario} \citep{2018MNRAS.476.4527T} and \textbf{emcee} \citep{foreman} to fit the WL 17 visibilities with this simple, 1-D geometrical model. The input visibilities were converted from a CASA visibility MS file to a uvtable with columns: u, v, Re, Im, w---where u and v are the visibility coordinates, Re is the real visibility data, Im is is the imaginary visibility data, and w is the weighting.

The model visibility computation was performed with \textbf{galario} by fitting a model to the radial intensity profile given by the visibilities listed in the uvtable. The image grid was then sampled to create an intensity map, and Fourier transformed to compute the real and imaginary visibilities at each point in this map. The $\chi^2$ was calculated for the total real and imaginary visibilities for the model data and observed data to find the best-fit parameters. We determined the number of walkers used in the MCMC ensemble sampler by taking 10 times the number of total dimensions. 

Convergence was achieved when all traces in a trace plot maintained a steady value for 500 steps or more. We selected our best-fit model to be the 50th quartile, or median, of the sampled range. We assessed the accuracy of a model by visually inspecting the uv plots and residual maps.

We ran the MCMC ensemble sampler with 1000 steps for the Gaussian ring model. Table \ref{tab:gausscav} gives the best-fit parameters for all three bands when fit separately.  We also fit the three bands simultaneously, leaving peak flux a free parameter and assuming a common disk shape but found that this does not improve the best-fit. Since the disk could exhibit wavelength-dependent properties, such as changes in optical depth, variations in the temperature structure, or differences in grain size, we continue analysis with the unfixed model which allows disk structure to vary between the three frequencies.

\begin{deluxetable*}{c|cccccc}[t!]
\tablecaption{Best-fit Model: Gaussian + Cavity \label{tab:gausscav}}
\tablehead{
\colhead{$\nu$} & \colhead{$f_0$} & \colhead{$f_p$} & \colhead{$\sigma$} & \colhead{$i$}  & \colhead{$\phi$} &  \colhead{$r_0$}\\
\colhead{(GHz)} & \colhead{(mJy)} & \colhead{(mJy/pix)} & \colhead{(mas)} & \colhead{(\degree)} & \colhead{(\degree)} & \colhead{(mas)}
}
\startdata
100 & 7.27$^{+0.05}_{-0.05}$ & 0.014 & 40.0$^{+1.4}_{-1.3}$ & 31.2$^{+1.7}_{-1.9}$ & 56$^{+4}_{-4}$ & 103.1$^{+1.7}_{-1.8}$ \\
233 & 50.16$^{+0.04}_{-0.04}$ & 0.089 & 40.0$^{+0.3}_{-0.3}$  & 33.10$^{+0.19}_{-0.18}$  & 63.1$^{+0.3}_{-0.3}$  & 111.9$^{+0.3}_{-0.3}$ \\
345  & 125.9$^{+0.3}_{-0.3}$ & 0.30 & 30.7$^{+0.6}_{-0.6}$  & 36.48$^{+0.5}_{-0.5}$  & 62.9$^{+1.0}_{-1.0}$  & 113.7$^{+0.6}_{-0.6}$ \\
\enddata
\tablecomments{The best-fit parameters for the Gaussian + cavity model applied separately to the 100 GHz, 233 GHz, and 345 GHz visibilities. $f_0$ is the total flux density of the model. See Equation (1) for an explanation of the other parameters. The errors correspond to the MCMC fit uncertainty only.}
\end{deluxetable*}

\begin{figure*}[th!]
\plotone{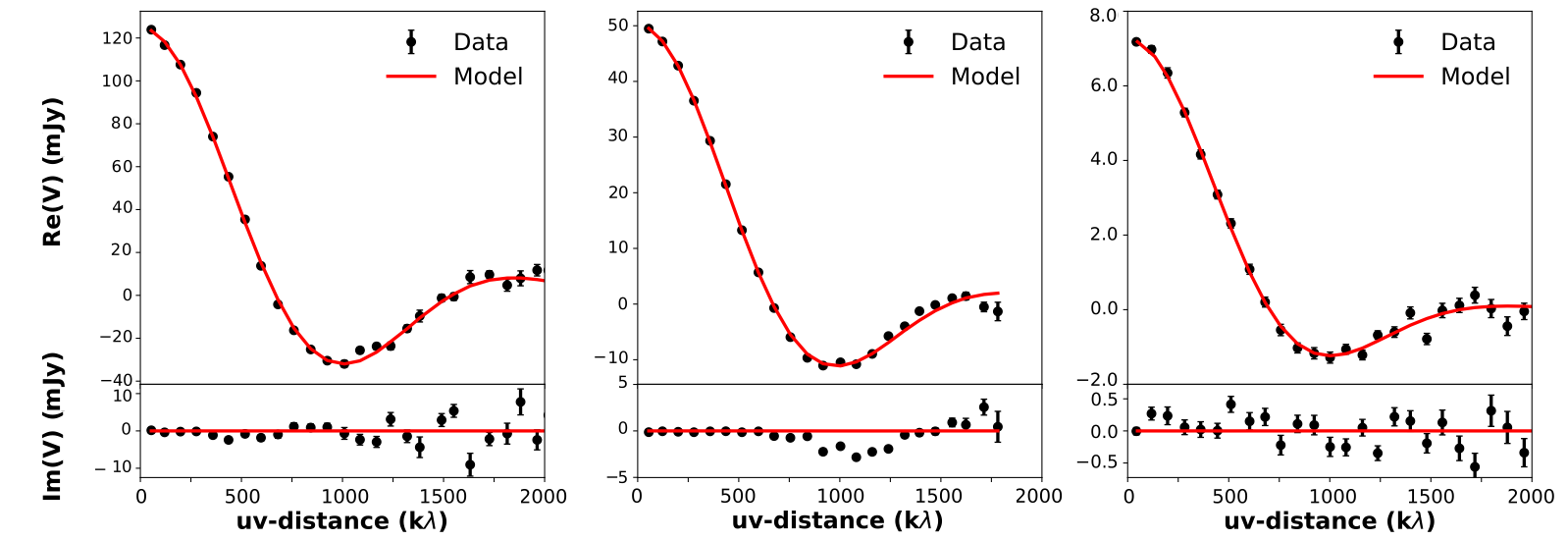}
\caption{The real and imaginary data shown with the best-fit Gaussian ring model plotted vs. uv-distance for WL 17 at 345 GHz (left), 233 GHz (middle), and 100 GHz (right). The real (top panel) and imaginary (bottom panel) data points are shown as black points, and the best-fit model as a red line.
\label{fig:all_final}}
\vspace{5mm}
\end{figure*}

 Figure \ref{fig:all_final} shows the visibility profiles at each band with the best-fit model in red.  In general, the Gaussian ring model provides a good fit to the real visibilities in all three bands.  There is some deviation in the 233 GHz data at the highest uv distances (> 1250 klambda).  Due to the very high sensitivity of the 233 GHz data, these deviations are well above the noise.  More striking, however, is that the 233 GHz data shows a significant non-zero imaginary component that is not seen in the lower sensitivity 345 GHz and 100 GHz data. The magnitude of this imaginary component is only about 8\% of the total flux. Considering the SNR of the 100 and 345 GHz data, and in particular the variations in the imaginary components compared to the total flux, such a deviation would likely remain undetectable if it is at the same level at those wavelengths. The non-zero imaginary component indicates that the emission in WL 17 is non-axisymmetric at least at 233 GHz.  We discuss this feature more in the next section.
 
 To try and account for the asymmetric emission in the 233 GHz data, we also fit the visibilities with a Nuker profile (see Appendix \ref{sec:nuker}).  We find that these models do not improve the overall fit and have reduced $\chi^2$ values that are  about a factor of 2 worse than the simple Gaussian ring model.  We note that the reduced $\chi^2$ is dominated by uncertainties in the real component so even though the Gaussian ring model fails to fit the imaginary visibilities, it provides the best fit to the real components. We include the model description and results for the Nuker profile in Appendix \ref{sec:nuker}.

 \subsection{Results from the Gaussian+cavity Model Fit}
 

Figure \ref{fig:robust} compares an image map of WL 17 at 233 GHz and the residual map obtained for the Gaussian ring model, both with robust = -2 to obtain the highest resolution. The residual map shows two bright peaks on either side of the disk, aligned with the major axis. These residual peaks are detected with S/N > 20, which means that these are significant features, but at a magnitude of only $\sim$4\% compared to the peak flux. We also see negative bowling of order 10$\sigma$ outside the residuals. This could be a contribution from the unfitted envelope component or due to radial variations in the disk physics which is not captured by a simple geometric model.



We re-fitted the 233 GHz data using a non-uniform ring model to represent the flux distribution from the ring and the double-lobed feature.  For simplicity, we assumed that the residuals could be represented by Gaussians with inclinations that were fixed to the inclination of the disk, but their position angles, sizes, and peak fluxes were allowed to vary.  We added these Gaussians to our original uniform Gaussian ring model at the position of each residual. We find that this three-component model did not fit the 233 GHz data as well as the simpler Gaussian+cavity model. The two Gaussians representing the residuals were poorly constrained, with very broad widths and large errors on flux.  More complex physical models will be necessary to accurately reproduce the double-lobed flux distribution at 233 GHz.  


We consider the double-lobed residuals to indicate a flux enhancement along the major axis at 233 GHz, although these features are not resolved in the 233 GHz beam.  First, the disk appears slightly asymmetric, with brighter emission in the north-east corner at both 345 GHz (see Figure \ref{fig:rp5}) and at 233 GHz with robust = -2 (see Figure \ref{fig:robust}).  This location of brighter emission matches well the location of the residuals, including the north-east residual being brighter.  Second, the visibility profile at 233 GHz shows a significant negative dip in the imaginary component of $\sim$-4 mJy around 1000 k$\lambda$.  This dip is at a uv-distance comparable to the ring cavity, which would imply non-axisymmetric emission in the ring itself.  These consistent elements indicate that the residuals are real features of the disk and not imaging artifacts.  We discuss possible origins of the double-lobed emission in Section \ref{sec:dlemission}.  
 

 
We also fitted the Gaussian+cavity model separately to the 345 GHz and 100 GHz data.  Figure \ref{fig:orgmodresid} shows the observations and the residuals for both bands.  We find a good fit to the observed visibilities with the simple Gaussian ring model, although the 345 GHz model appears to overestimate the flux in the ring and the 100 GHz model appears to underestimate the flux in the ring.  Nevertheless, the residuals at 345 GHz are all below 5$\sigma$, whereas the 100 GHz data show a residual of ~5$\sigma$ over half of the disk. We note that the residuals at 100 GHz do not align with the lobes seen at 233 GHz and are likely due to an underestimate of the total flux by the Gaussian ring model. The suggested asymmetry in the 345 GHz image could not be confirmed from our visibility analysis and might be the result of imaging artefacts.

\renewcommand\thefigure{\arabic{figure}}
\begin{figure*}[t!]
\plotone{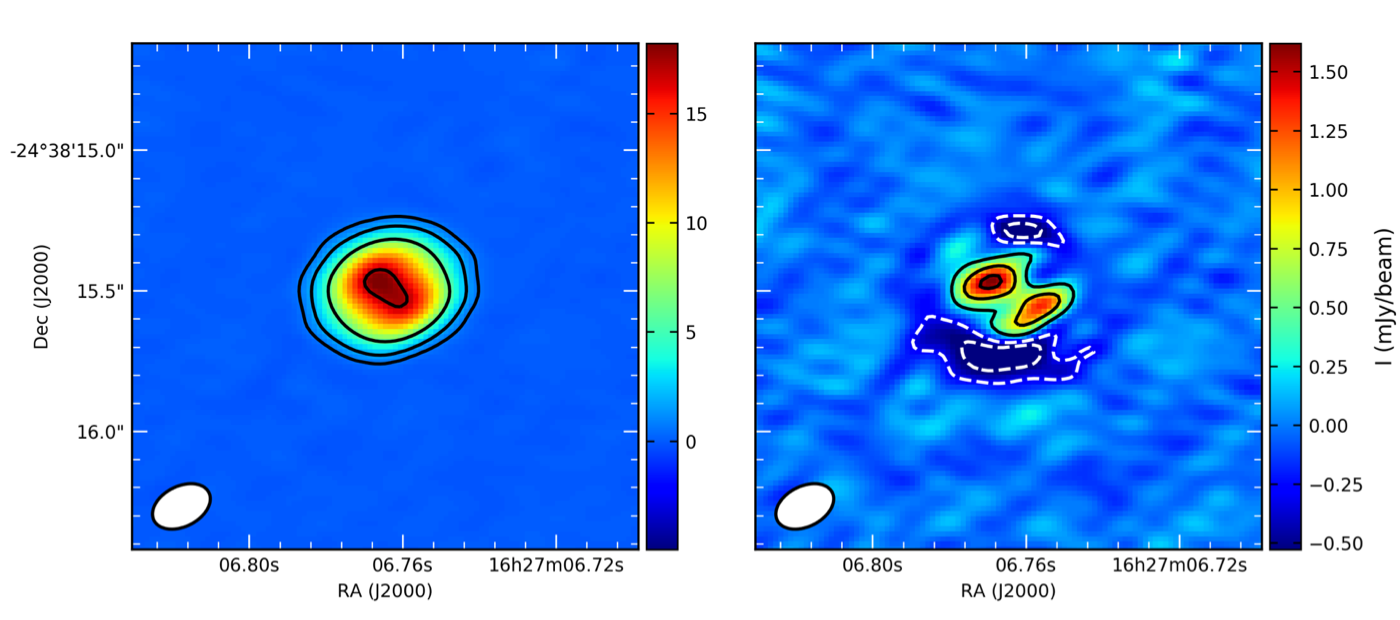}
\caption{Double-Lobe emission at 233 GHz. Left: WL 17 imaged with robust = -2. Positive contours correspond to 30$\sigma$, 90$\sigma$, 270$\sigma$, and 320$\sigma$.  Right: The residual image of WL 17 with robust = -2.  The residual is constructed by subtracting the Gaussian ring model visibilities from the original WL 17 visibilities. Positive (solid) contours start at 5$\sigma$ and increase by 10$\sigma$, and negative (dashed) contours start at  -5$\sigma$ and decrease by 5$\sigma$. Both left and right images have $\sigma$ = 63 $\mu$Jy/beam and a beam size of 0.22" by 0.14". The residual image shows a double-lobed feature, indicating additional substructure in the disk. The location of the double-lobed feature is in agreement with the north-east brightness asymmetry seen in Figure \ref{fig:rp5}. \label{fig:robust}}
\end{figure*}

\begin{figure*}[th!]
\plotone{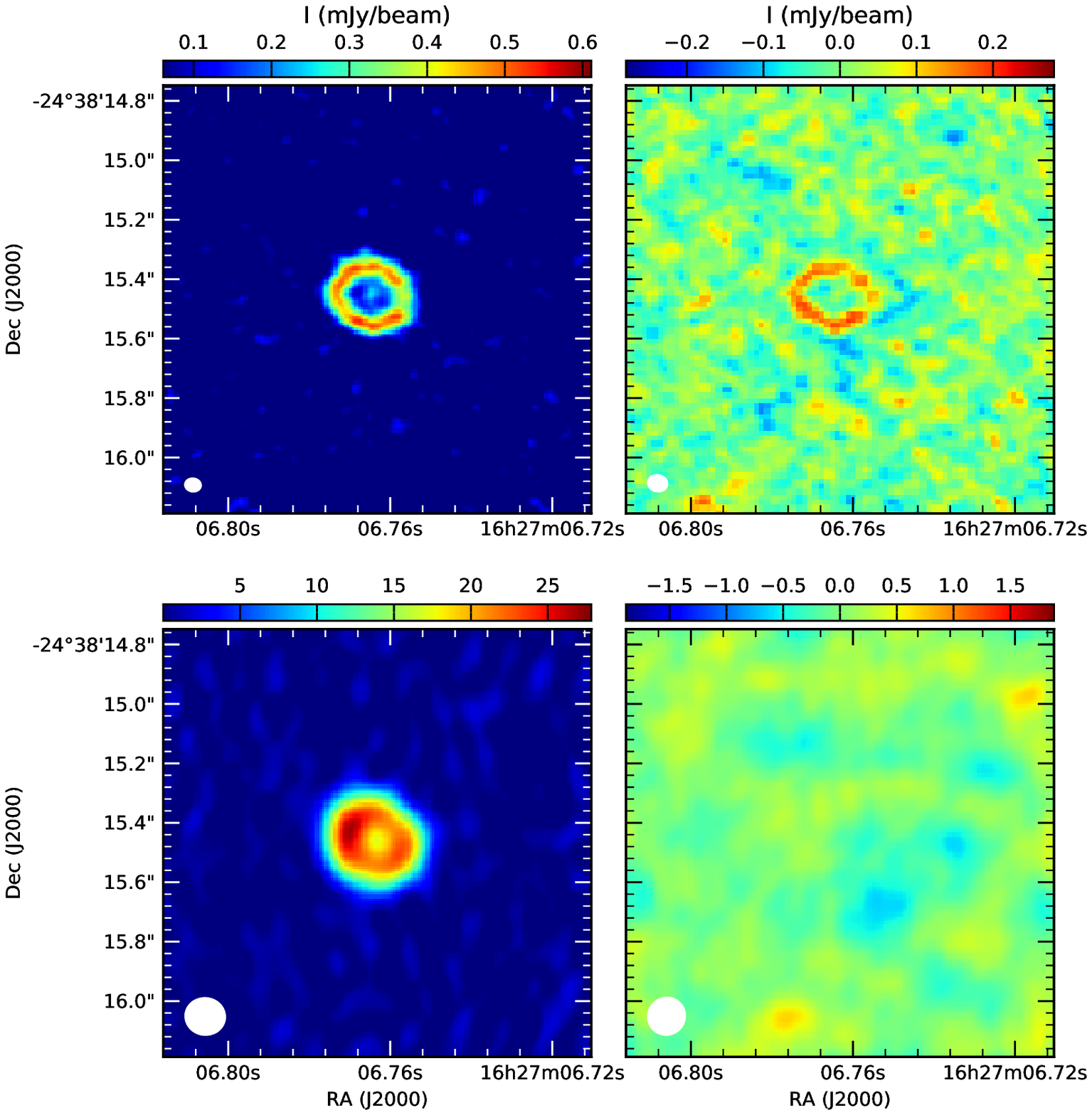}
\caption{The original and residual images for 100 GHz (top) and 345 GHz (bottom). The 100 GHz data are cleaned with robust = 0.5 and the 345 GHz data are  cleaned with robust = -2. The residual maps correspond to the original image minus the Gaussian ring best-fit model. The colorbar for both residual images shows $\pm7\sigma$, where $\sigma_{100} = 35\ \mu$Jy/beam and $\sigma_{345} = 730\ \mu$Jy/beam.
\label{fig:orgmodresid}}
\end{figure*}
 
 Ultimately, higher resolution observations are necessary to apply more complex disk models to WL 17.  Since the double-lobed feature is 40 times fainter than the peak emission in the disk, we will continue our analysis using the Gaussian ring model.

\newpage

\section{Discussion}\label{sec:discuss}

\subsection{Estimating Optical Depth}\label{sec:opdepth}

Since WL 17 is a bright, large disk, we would expect it to display polarization from dust self-scattering. Nevertheless, WL 17 is undetected in dust polarization \citep{sarah}. An optically thick disk is more likely to exhibit dust self-scattering, therefore we measure the optical depth of WL 17 at all three frequencies to determine whether WL 17's optical depth is consistent with a lack of self-scattering.

To estimate the optical depth of WL 17, we model the specific intensities of our best-fit Gaussian ring models for the 100 GHz, 233 GHz, and 345 GHz data using:

\begin{equation}
I(\nu) = B(\nu, T)(1-e^{-\tau_{\nu}})
\end{equation}

\noindent where $B(\nu, T)$ is the standard black body function and $\tau_{\nu} = \tau_{233}\left(\frac{\nu}{233 GHz}\right)^\beta$ is the optical depth of a given frequency scaled by the 233 GHz data. To find the best-fit values for $\tau_{233}$ and $\beta$ for assumed temperatures, we compute the model intensity at each radius using the best-fit Gaussian ring model and use an MCMC algorithm to fit the intensities for the three frequencies simultaneously. We adopt two different temperature profiles. First, we use the temperature distribution obtained by \citet{sheehan} in their radiative transfer model of WL 17.  Second, we use the midplane temperature profile for a flared disk adopted from \citet{dullemond2} and \citet{chiang}:

\begin{equation}
T(r) = \left(\frac{\phi L_{*}}{8 \pi \sigma_{SB} r^2}\right)^{1/4}
\end{equation}

\begin{figure}[th!]
\includegraphics[width=85mm]{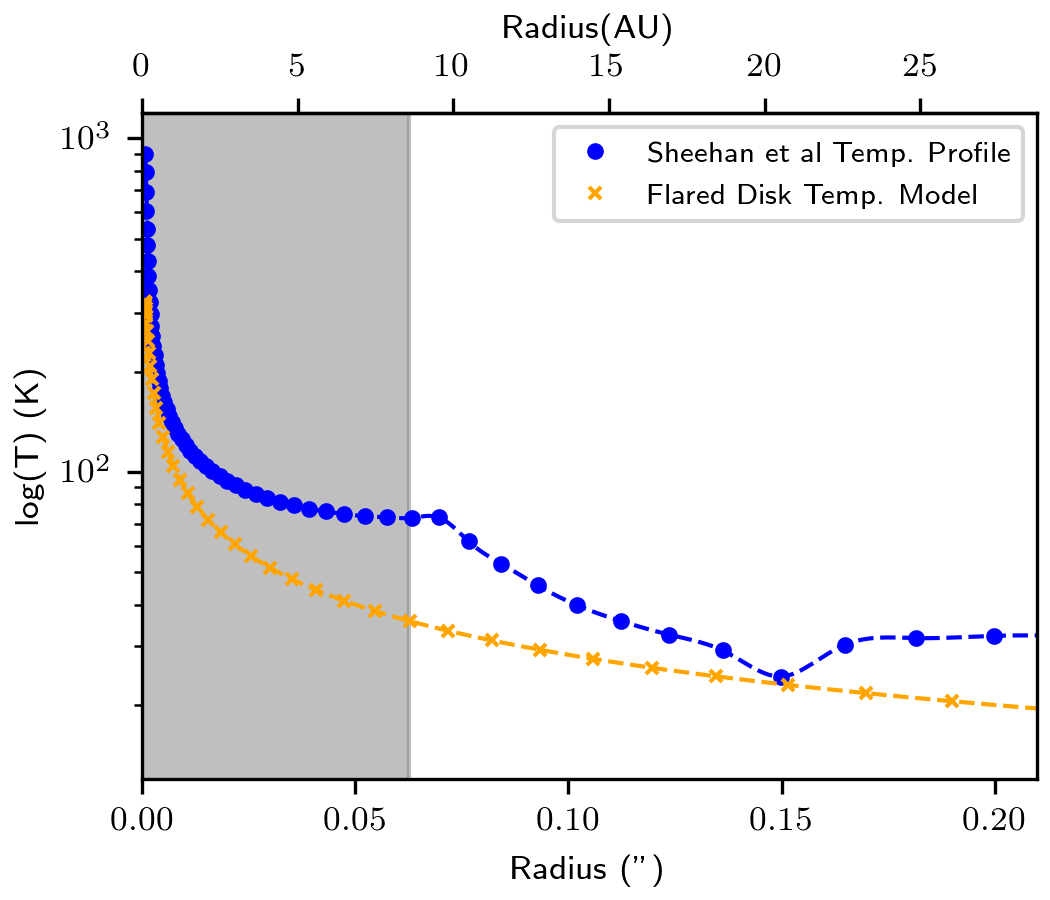}
\caption{Temperature versus radius in WL 17. The two temperature models used in Section \ref{sec:opdepth} are plotted as a function of radius, where blue points corresponds to the the radiative transfer model temperatures from \citet{sheehan} and orange points corresponds to the flared disk model from \citet{dullemond2, chiang}. The dashed lines follow a 2D interpolation between respective points.
\label{fig:temp}}
\end{figure}

\noindent Figure \ref{fig:temp} compares the adopted temperature profiles for our disk. The warmer temperatures in the \citet{sheehan} radiative transfer temperature model likely arise from the effects of their included envelope component, which insulates and heats the disk \citep{dalessio, vant}.

\begin{figure}[h!]
\includegraphics[]{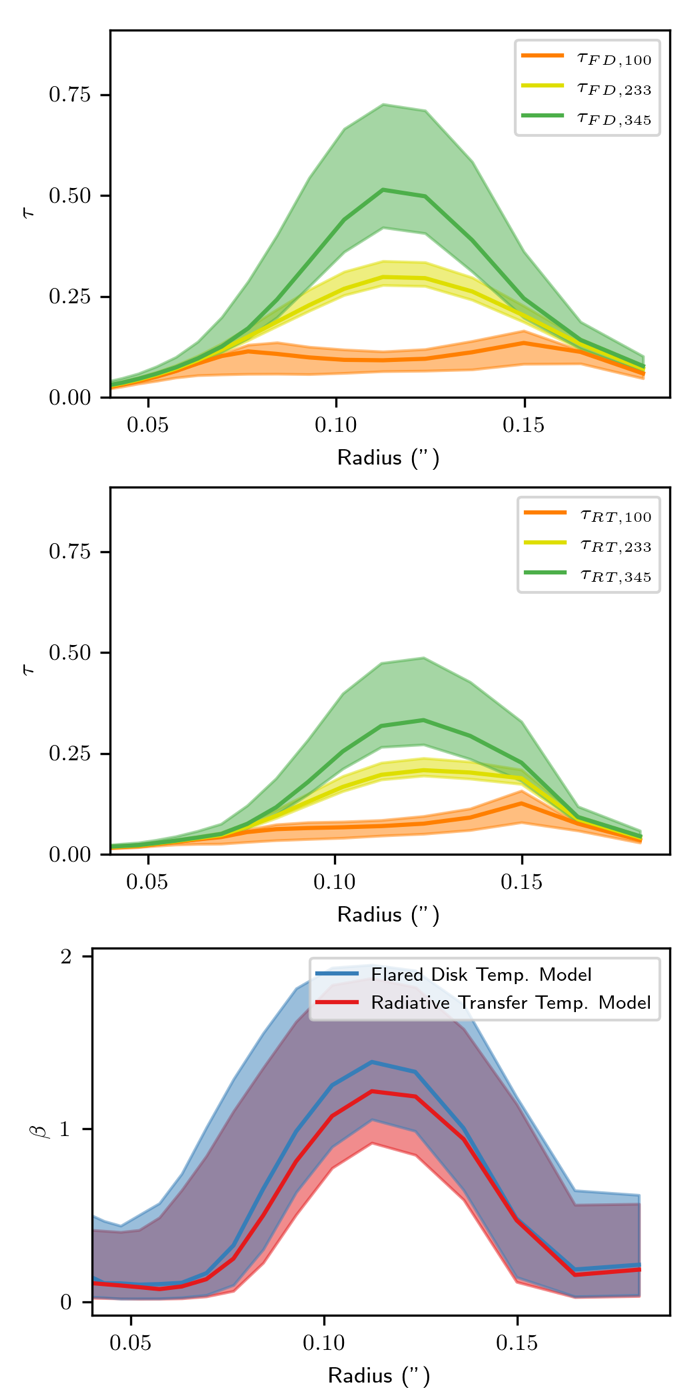}
\caption{The optical depth for the 100 GHz, 233 GHz, and 345 GHz data calculated with the flared disk model from \citet{dullemond2} and \citet{chiang} (top), the optical depth calculated with the radiative transfer temperature model from \citet{sheehan} (middle), and the $\beta$ values for each temperature model (bottom). The solid lines show the median sampled values, and the shaded areas show the range of accepted values within a 5$\%$ flux calibration error, the fit error, and the map RMS uncertainties. \\
\hfill \\
\newpage
\label{fig:specind1}}
\end{figure}

Figure \ref{fig:specind1} shows $\tau_{233}$ and $\beta$ as a function of radius for both temperature profiles. The dashed and dotted lines trace the best-fit profile while light colored regions around the line show a range of values selected within the 5$\%$ flux calibration error, and the fit and map RMS uncertainties. The light colored region is generated by finding $\tau_{233}$ and $\beta$ at each radius for 100 randomly modified intensity maps within the given uncertainties. The intensity maps are calculated for each frequency using the Gaussian ring model, where $f_0$, $\sigma$, and $r_0$ are randomly sampled from a normal distribution of values within each parameter's uncertainties. The plot shows a maximum optical depth for the 233 GHz data ranging from $\sim$0.27 to 0.35 for the flared disk model and from $\sim$0.17 to 0.25 for the temperature profile from \citet{sheehan}. We find that the 345 GHz data are more optically thick, with a maximum optical depth ranging from $\sim$0.4--0.77 for the flared disk temperature model and $\sim$0.3--0.5 for the radiative transfer temperature model. The 100 GHz data are very optically thin with a maximum optical depth of $\sim$0.075--0.175 for both temperature models.

The 100 GHz optical depth may be more affected by the underestimation of the model fluxes. For example, Figure \ref{fig:orgmodresid} shows nearly uniform residuals of $\sim 0.2$ mJy beam$^{-1}$ around the ring.  This missing emission  could cause the deviation from a Gaussian-like $\tau$ distribution in Figure \ref{fig:specind1}. We fitted the 100 GHz data over different uv ranges to test whether or not the noisy, higher uv-distances affected the fit, but found consistent best-fit results. These results suggest that the disk at 100 GHz is non-Gaussian in shape.  We attempted to fit the disk with a Nuker profile, however, the sensitivity was insufficient to constrain those models (see Appendix \ref{sec:nuker}). Underestimated emission at 100 GHz could result in lower $\tau$ values toward the peak of the ring as shown by the dip in Figure \ref{fig:specind1}, which could also mean that $\beta$ is slightly overestimated. While a flatter $\beta$ value could result in the 345 GHz data approaching $\tau = 1$, we expect that the 233 GHz and 100 GHz emission would still remain optically thin.  

These values indicate that the WL 17's disk is optically thin at all radii at 233 GHz, which could explain why it was undetected in dust polarization from self-scattering in \citet{sarah}. Additionally, WL 17 is fairly face on (i $\sim$ 30$\degree$), which could explain why it appears optically thin at 233 GHz even with its relatively large dust mass compared to other disks in Ophiuchus. The 345 GHz band have $\tau$ closer to 1. Future studies could use this band to test whether or not dust self-scattering is present in this disk.

\subsection{Multi-Wavelength Disk Comparison}\label{sec:wavcomp}

The 233 GHz data show a strong ($>$ 20$\sigma$) feature in the imaginary profile. 
This dip reveals a feature in the azimuthal versus radial intensities caused by substructures in the disk at approximately the same spatial scales as the cavity radius.

The 100 GHz and 345 GHz data do not show a negative dip in their imaginary visibilities around 1000 k$\lambda$, thus the Gaussian ring model provides a better overall fit to their visibilities (see Figure \ref{fig:orgmodresid}). However, a deviation in the imaginary visibilities with a similar magnitude to that found in the 233 GHz data cannot be ruled out due to the lower sensitivities of the 100 GHz and 345 GHz data. The 233 GHz data have a point source sensitivity that is a factor of 3-8 times better than the 345 GHz and 100 GHz data assuming $\beta$ = 0 and T = 50 K. Even though Figure \ref{fig:rp5} and Figure \ref{fig:orgmodresid} show a clear asymmetry in the 345 GHz data, the asymmetry is small relative to the map sensitivities.


The higher sensitivity of the 233 GHz data allows us to fit more complex models to its real and imaginary visibilities. Even so, when fitting the 233 GHz data with a Nuker profile, which provides more flexibility when fitting multiple disks with different profile shapes \citep{powell}, the $\alpha$ parameter describing the inner transition disk is unconstrained (see Appendix \ref{sec:nuker}). We try to fit the higher resolution 100 GHz data with the Nuker profile to help constrain $\alpha$, but find the model will not converge because the 3 mm emission is too faint to constrain the $\beta$ and $\gamma$ parameters. We attempt also to fit a Nuker profile to the 233 GHz and 100 GHz data simultaneously, but even the combined data are unable to converge on a best-fit model. Therefore, the single Gaussian ring is the best converging model for these three bands at the current resolutions and sensitivities.

\subsection{Double-Lobe Emission}\label{sec:dlemission}

A key feature in the 233 GHz data is the previously undetected double-lobed feature seen in the Stokes I continuum data. One-sided lobes have been seen in several protoplanetary disks, e.g. \citep{jane, miley, stammler, shen, kanagawa, nienke}. Such features show a clear wavelength dependence, as they are generally more pronounced at longer wavelengths \citep{vandermarel2015,cazzoletti2018}. A dependence could not be measured for WL17 due to the difference in the three frequencies' sensitivities.

These lobes have been attributed to dust or pressure traps caused by inhomogenetities in the gas density profile of a disk \citep{birnstiel}, chemical variations in the disk that affect the dust density or size distribution \citep{hein}, dust accumulation in gravitational wells around by a nearby planet \citep{jane, Baruteau, stammler}, or geometric effects like inclination of a geometrically thick, flared disk \citet{doi}.

The location of the features of nearly equal magnitude on either side of the major axis suggests that the 
lobes could be due to geometric effects.  For example, if the disk is geometrically thick and flared, the low inclination could result in a slightly larger column of dust along the major axis compared to the minor axis. Since the optical depth at 233 GHz is estimated at $\tau$ < 0.3, a slight difference in dust column would translate to an increase in flux.  If the dust emission at 233 GHz is coming from a geometrically thick disk, then the true disk profile of this source will require more complex models that are beyond the scope of the current paper.

Indeed, \citet{doi} recently modeled the dust emission from geometrically thick, inclined rings that have large scale heights. They found that optically thin rings should show a double-lobed feature, with the brighter emission associated with the lobes appearing on either side of the rings' major axes.  They also showed that the excess emission is weaker or not detectable for disks that are optically thick with $\tau \gtrsim$ 1. Their analysis agrees well with our observations at 233 GHz and 345 GHz.  Both bands appear optically thin, with $\tau_{233} \sim 0.25$ and $\tau_{345} \sim 0.5-0.7$, and both bands show at least one brighter lobe along the major axis.  The 233 GHz data show the full double-lobed profile expected from \citet{doi}, whereas the 345 GHz data show only a single brighter lobe at the northern edge of the major axis.  The 345 GHz are likely not sensitive enough to detect the fainter southern lobe. 
The 100 GHz data is the most optically thin with $\tau$ < 0.1, however, we are unable to detect the lobes in this disk either due to a lack of sensitivity or because the dust emission at 3 mm is more symmetric.

We propose that our observations at 233 GHz indicate that the 1.3 mm dust emission is not tracing a thin midplane, but a flared dust disk. This difference could affect both models of dust polarization as well as models of the temperature structure observed by ALMA.  We cannot estimate the scale height at 233 GHz at this time, however.  The models of \citet{doi} require very high angular resolution to disentangle the resolution of the disk from the geometric disk features.  Our data at 233 GHz do not spatially resolve the ring, although with our high sensitivity we are still able to fit the ring visibilities well.  Higher resolution observations taken with high sensitivity will be necessary to capture the double-lobed feature well enough to estimate a scale height for the disk.



\section{Conclusions}\label{sec:conc}

We fit the WL 17 ALMA data at 345 GHz, 233 GHz, and 100 GHz with simple geometric models to infer disk intensity profiles and measure the optical depth through the disk.  Our main conclusions are:

\begin{enumerate}
  \item The disk emission is generally consistent with a simple Gaussian ring model.
  \item The estimated optical depth at 233 GHz is $\tau$ < $\sim$0.3, which is indicative of optically thin dust emission. The low optical depth could explain why WL 17 is undetected in polarization from dust self-scattering, even though it is a bright and relatively massive disk. The 345 GHz data is closer to $\tau$ $\sim$1, which makes it better for dust self-scattering observations.
  \item The high sensitivity of the 233 GHz data revealed a previously undetected double-lobed feature along the major axis. Higher resolution observations are needed to fit this substructure.
  \item We propose that the double-lobed feature is caused by a flared dusk disk that has a slightly higher column of dust along the major axis over the minor axis.  More complex disk geometries will be necessary to model the disk and predict its polarization patterns from self-scattering.
\end{enumerate}

The double-lobed feature seen in WL 17 at 233 GHz was detected because dust polarization measurements with ALMA need longer integration times on sources.  Such features may be present in other disks that have been studied in dust polarization and offer an opportunity to study the structure of disks in greater detail than most continuum-only studies are capable of doing.

\section{Acknowledgements}

The authors wish to thank the anonymous referee for their helpful comments that improved the discussion of the paper. The authors would also like to thank A. Kataoka for valuable discussion about the double-lobed emission detected in WL 17. HG acknowledges support from the SAO REU program, which is funded in part by the National Science Foundation REU and Department of Defense ASSURE programs under NSF Grant no.\ AST-1852268, and by the Smithsonian Institution. SIS acknowledges support from the Natural Science and Engineering Research Council of Canada (NSERC), RGPIN-2020-03981. N.M. acknowledges support from the Banting Postdoctoral Fellowships program, administered by the Government of Canada. This paper makes use of the following ALMA data: ADS/JAO.ALMA$\#$2015.1.01112.S and  ADS/JAO.AL- MA$\#$2015.1.00761.S. ALMA is a partnership of ESO (representing its member states), NSF (USA) and NINS (Japan), together with NRC (Canada), MOST and ASIAA (Taiwan), and KASI (Republic of Korea), in cooperation with the Republic of Chile. The Joint ALMA Observatory is operated by ESO, AUI/NRAO and NAOJ. The National Radio Astronomy Observatory is a facility of the National Science Foundation operated under cooperative agreement by Associated Universities, Inc.

\appendix
\section{Nuker Profile} \label{sec:nuker}

The Nuker profile has been used more recently to fit protoplanetary disks and offers more flexibility than a Gaussian profile \citep{lauer}. We fit a Nuker profile to the 233 GHz data in \textbf{galario} using the following equation:

\begin{equation}
I(r) = f_0 \left(\frac{r}{R_t}\right)^{-\gamma}\left[1+ \left(\frac{r}{R_t}\right)^\alpha\right]^{(\gamma -\beta)/\alpha}
\end{equation}

\noindent where $f_0$ is the normalization parameter, $R_t$ is the transition radius, $\gamma$ is the inner disk index, $\beta$ is the outer disk index, and $\alpha$ is the transition disk index. The $\alpha$ parameter describes the asymptotic behavior of the disk. We found that the 233 GHz data alone did not have sufficient resolution to constrain $\alpha$ well.  We therefore fit the profile fixing $\alpha$ = 1.0. We include a position angle, inclination, and offset parameters for a total of four geometric and four physical parameters: $f_0$, $R_t$, $\gamma$, $\beta$, $\phi$, $i$, $d \alpha$, and $d\delta$

We run the \textbf{emcee} ensemble sampler for 5000 iterations. Table \ref{tab:nuker} shows the minimum and maximum value for the sampled parameter space, the initial guess $p_0$, the computed best-fit, and the corresponding best-fit errors. Figure \ref{fig:nuker} shows the best-fit plotted with the real and imaginary visibilities as a function of uv-distance.

The Nuker profile provides a good fit to the real visibilities, only diverging from the data at uv-distances higher than $\approx$ 1600 k$\lambda$. However, the Nuker profile fails to fit the imaginary visibilities. Like the Gaussian ring model, the Nuker profile is symmetric and therefore does not fit the imaginary data well.  There is slight structure in the imaginary component of the Nuker model in Figure \ref{fig:nuker}, but we attribute this to limitations in our resolutions.  The WL 17 disk width is is smaller than than our limiting resolution.  A residual image between the original 233 GHz data and the Nuker profile is identical to the residual shown in Figure \ref{fig:robust} for the Gaussian ring model.

The higher sensitivity of the 233 GHz data allows us to fit more complex models to its real and imaginary visibilities. Even so, when fitting the 233 GHz data with the Nuker profile the $\alpha$ parameter is unconstrained. To improve the constraint on alpha, we also fit a Nuker profile to the 233 GHz and the higher resolution 100 GHz data simultaneously. Nevertheless, we were unable to converge on a best-fit model. The 100 GHz data lack the sensitivity to constrain beta and gamma, and the disk profiles may be different for the two bands due to differences in dust grain distributions, temperature gradients, and optical depth. Therefore, the single Gaussian ring is the best converging model for our analysis.

\begin{centering}
\centering
\begin{figure*}[th!]
\includegraphics[width=80mm]{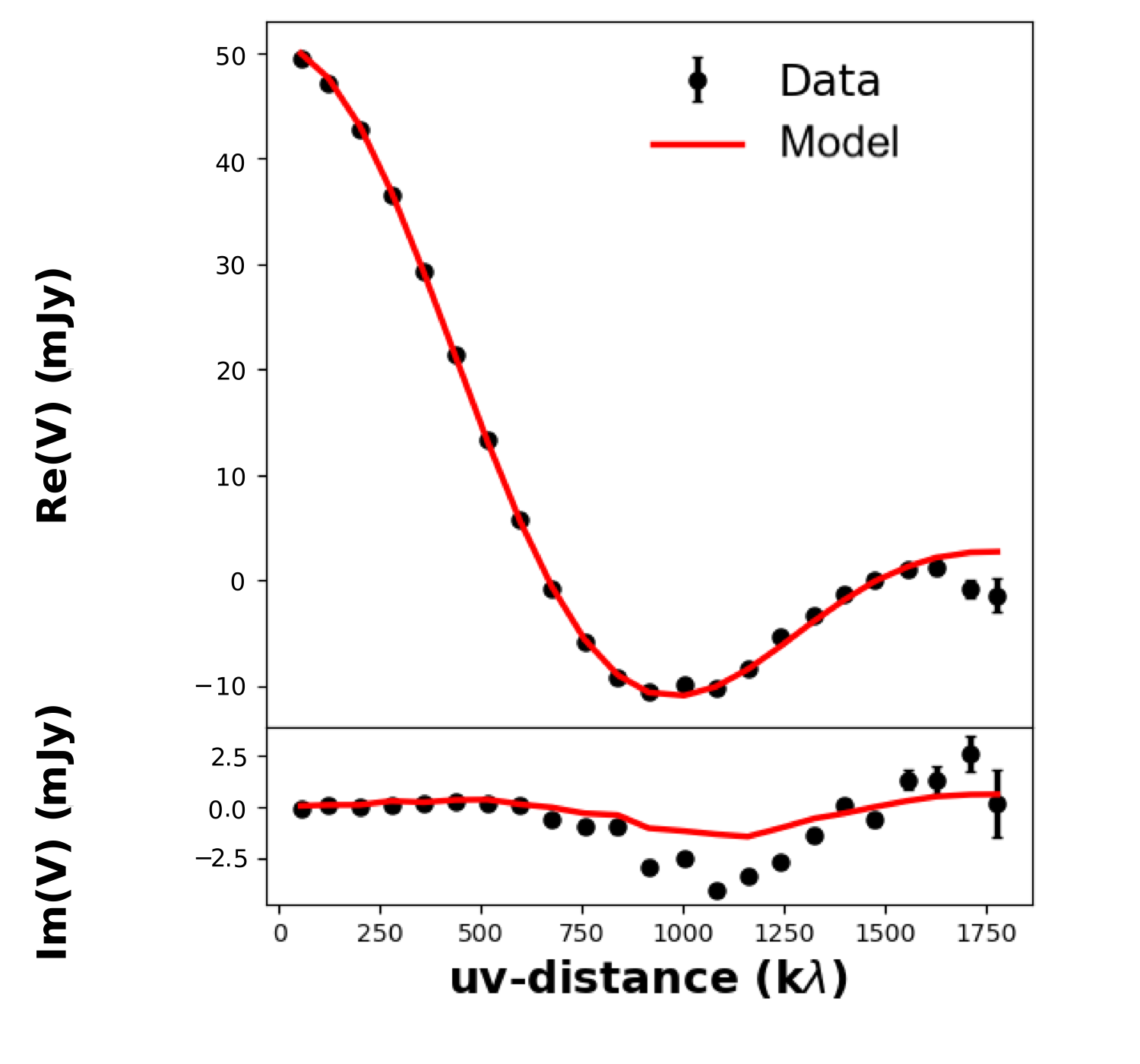}
\caption{The real and imaginary visibilities as a function of uv-distance for WL 17 with the best-fit model from the Nuker profile. The original data is shown as the black points, and the best-fit is shown as a red line.
\label{fig:nuker}}
\end{figure*}
\end{centering}

\begin{deluxetable}{c|ccccc}[b!]
\tablecaption{Best-fit Model: Nuker Profile \label{tab:nuker}}
\tablehead{
\colhead{} & \colhead{min} & \colhead{max} & \colhead{$p_0$}& \colhead{best-fit}& \colhead{errors}
}
\startdata
$i$ & 0 & 90 & 50 & 33.06 & $\pm^{0.2102}_{0.1895}$\\
log$_{10}f_0$ & -3 & 0 & -1.29 & -1.30 & $\pm^{0.0012}_{0.0012}$ \\
$\phi_{disk}$ & 20 & 100 & 64 & 64.15 & $\pm^{0.3603}_{0.3353}$\\
$R_t$ & 0.0001 & 10 & 0.14 & 0.15 & $\pm^{0.0008}_{0.0010}$ \\
$\gamma$ & -5 & 2 & -1.5 & -1.49 & $\pm^{0.0432}_{0.0450}$ \\
$\beta$ & 2 & 14 & 9 & 13.72 & $\pm^{0.2011}_{0.4127}$\\
\enddata
\tablecomments{The minimum and maximum sampled range values, initial guess, MCMC ensemble sampler calculated best-fit (see Section \ref{sec:obs}, and best-fit errors for the Nuker profile parameters.}
\end{deluxetable}

\end{document}